\documentclass[prl,twocolumn,showpacs,superscriptaddress]{revtex4-1}
\usepackage{graphicx}
\usepackage{mathrsfs}
\usepackage{amsmath}
\usepackage{amssymb}
\usepackage{amsfonts}
\usepackage[normalem]{ulem}
\usepackage{color}
\newcommand{\tr}{\textcolor{red}}

\newcommand{\bra}[1]{\langle #1 | \,}
\newcommand{\ket}[1]{\, | #1 \rangle}
\newcommand{\braket}[2]{\langle #1 | #2 \rangle}
\newcommand{\expv}[1]{\left\langle #1 \right\rangle}
\newcommand{\ga}{\ga}
\newcommand{\Ga}{\Gamma}
\newcommand{\la}{\lambda}
\newcommand{\De}{\Delta}
\newcommand{\ve}{\varepsilon}
\newcommand{\us}{\uparrow}
\newcommand{\ds}{\downarrow}
\newcommand{\bk}{b^{\dagger}}
\newcommand{\cdk}{c^{\dagger}}
\newcommand{\ck}{c}
\newcommand{\bl}{\begin{linenomath*}}
\newcommand{\el}{\end{linenomath*}}
\newcommand{\E}{\hat{\mathcal{E}}}
\renewcommand{\P}{\hat{\mathcal{P}}}
\newcommand{\mS}{\hat{\mathcal{S}}}
\newcommand{\mSd}{\hat{\mathcal{S}}^\dagger}
\renewcommand{\S}[1]{\hat\sigma_{#1}}
\newcommand{\s}{\sigma}
\newcommand{\D}{\hat\Psi}
\newcommand{\B}{\hat\Phi}
\newcommand{\Dd}{\hat\Psi^\dagger}
\newcommand{\Bd}{\hat\Phi^\dagger}
\newcommand{\e}{\text{e}}
\newcommand{\den}[2]{{}_{#1}\!\rho_{#2}}

\newcommand{\st}{\text{s}_\theta}
\newcommand{\ct}{\text{c}_\theta}

\newcommand{\pdz}{\frac{\partial}{\partial z}}
\newcommand{\pdzz}{\frac{\partial^2}{\partial z^2}}
\newcommand{\pdx}{\frac{\partial}{\partial x}}
\newcommand{\pdt}{\frac{\partial}{\partial t}}
\newcommand{\pdr}{\frac{\partial}{\partial r}}
\newcommand{\pdR}{\frac{\partial}{\partial R}}
\newcommand{\pdrr}{\frac{\partial^2}{\partial r^2}}
\newcommand{\bea}{\begin{eqnarray}}
\newcommand{\eea}{\end{eqnarray}}
\renewcommand{\vec}[1]{\mathbf{#1}}
\renewcommand{\sp}{\hat\sigma^+}
\newcommand{\sm}{\hat\sigma^-}
\newcommand{\sz}{\hat\sigma^z}
\renewcommand{\a}{\hat a}
\newcommand{\ad}{\hat a^\dagger}
\newcommand{\de}{\hat d}
\newcommand{\ded}{\hat d^\dagger}
\newcommand{\ce}{\hat c}
\newcommand{\ced}{\hat c^\dagger}
\newcommand{\be}{\hat b}
\newcommand{\bed}{\hat b^\dagger}
\renewcommand{\ga}{\hat\gamma}
\newcommand{\gad}{\hat \gamma^\dagger}
\newcommand{\n}{\hat n}
\newcommand{\m}{\hat m}

\usepackage{hyperref}

\newcommand{\todo}[1]{{\bfseries\textcolor{red}{TO-DO: #1}}}
\newcommand{\mm}[1]{{\textcolor{blue}{#1}}}

\usepackage{cancel,ifthen}
\usepackage[normalem]{ulem}

\newcommand{\comment}[2][NoInPuT]{\ifthenelse{\equal{#1}{NoInPuT}}{}{{\color{blue}\sout{#1}}}{\color{red} #2}}
\renewcommand{\BibitemShut}[1]{}
\begin{document}

\title{Dissipative Preparation of Spatial Order in Rydberg-Dressed \\ Bose-Einstein Condensates}

\author{Johannes Otterbach}
\email{jotterbach@physics.harvard.edu}
\affiliation{Physics Department, Harvard University, 17 Oxford Street, Cambridge, MA 02138, USA} %

\author{Mikhail Lemeshko} 
\email{mikhail.lemeshko@gmail.com}
\affiliation{ITAMP, Harvard-Smithsonian Center for Astrophysics, 60 Garden Street, Cambridge, MA 02138, USA}%
\affiliation{Physics Department, Harvard University, 17 Oxford Street, Cambridge, MA 02138, USA} %

\date{\today}


\begin{abstract}
We propose a technique for engineering momentum-dependent dissipation in Bose-Einstein condensates with non-local interactions. The scheme relies on the use of momentum-dependent dark-states in close analogy to velocity-selective coherent population trapping. During the short-time dissipative dynamics, the system is driven into a particular finite-momentum phonon mode, which in real space corresponds to an ordered structure with non-local  density-density correlations. Dissipation-induced ordering can be observed and studied in present-day experiments using cold atoms with dipole-dipole or off-resonant Rydberg interactions. Due to its dissipative nature, the ordering does not require artificial breaking of translational symmetry by an optical lattice or harmonic trap. This opens up a perspective of direct cooling of quantum gases into strongly-interacting phases.
\end{abstract}

\maketitle
In condensed matter, structures with non-local correlations, such as crystals or liquids, arise due to conservative Coulomb forces between electrons and nuclei, giving rise to equilibrium configurations with some type of spatial order \cite{Solyom2007, LiebRMP76}. At sufficiently low temperatures, such structures can be treated as closed quantum systems, whose properties are encoded in the ground state of a corresponding Hamiltonian.

However, if a polyatomic quantum system is open, i.e. coupled to a fluctuating environment, its properties can be fundamentally altered. In a number of applications, such as quantum information processing \cite{NielsenQIP10} and coherent spectroscopy \cite{DemtroderLaserSpec}, environment-induced dissipation leads to undesired decoherence. In some other cases, interaction with the environment can lead to novel effects, such as the localization transition in the spin-boson model~\cite{LeggettRMP87}, or enhanced efficiency of energy transfer in photo-synthetic complexes \cite{LambertNatPhys12}. Understanding effects of the environment represents a formidable challenge due to the inherent complexity of `conventional' condensed matter systems and uncontrollable character of dissipation. However, recent experimental progress in the quantum control of cold atoms, trapped ions, optical and microwave photons, and defects in solids makes such an understanding more tractable \cite{NatQantSim12, LaddNatPhys10, 
JelezkoNJP12}. This allows the simulation of many-body Hamiltonians with controllable couplings to a dissipative environment \cite{NatQantSim12, MullerAdv12}.

Furthermore, this degree of controllability can be used to turn dissipation into a resource for quantum state preparation as has been shown theoretically \cite{Diehl2008,Verstraete2009,Weimer2010,Diehl2010,DallaTorre2013,LeePRL13}. The essence of this approach lies in a suitable engineering of the system-to-reservoir couplings, such that the dissipative time evolution drives the system towards an interesting steady state. Using dissipation for quantum state preparation has been demonstrated in experiments with trapped ions \cite{Barreiro2011}.

Recently Lemeshko and Weimer showed that dissipation-induced non-conservative forces between ultracold atoms can result in the formation of `dissipatively-bound molecules' \cite{LemWeimDiss, LemFPCCP13}. As a step forward, we demonstrate here that dissipation can induce spatial ordering in a many-body system.  As opposed to `conventional' condensed matter systems whose spatial order occurs due to spontaneous symmetry breaking in a corresponding Hamiltonian, the  structures discussed here appear as a transient in the short-time dynamics of a driven-dissipative continuum system, due to the competition between coherent and incoherent time-evolution. Unlike for dipolar crystals self-assembled in a harmonic trap \cite{BuchlerPRL07}, dissipatively ordered structures can be realized in free space or a uniform trap \cite{GauntPRL13}, i.e. without artificially breaking the translational symmetry.
\begin{figure}[b]
  \includegraphics[width=.75\linewidth]{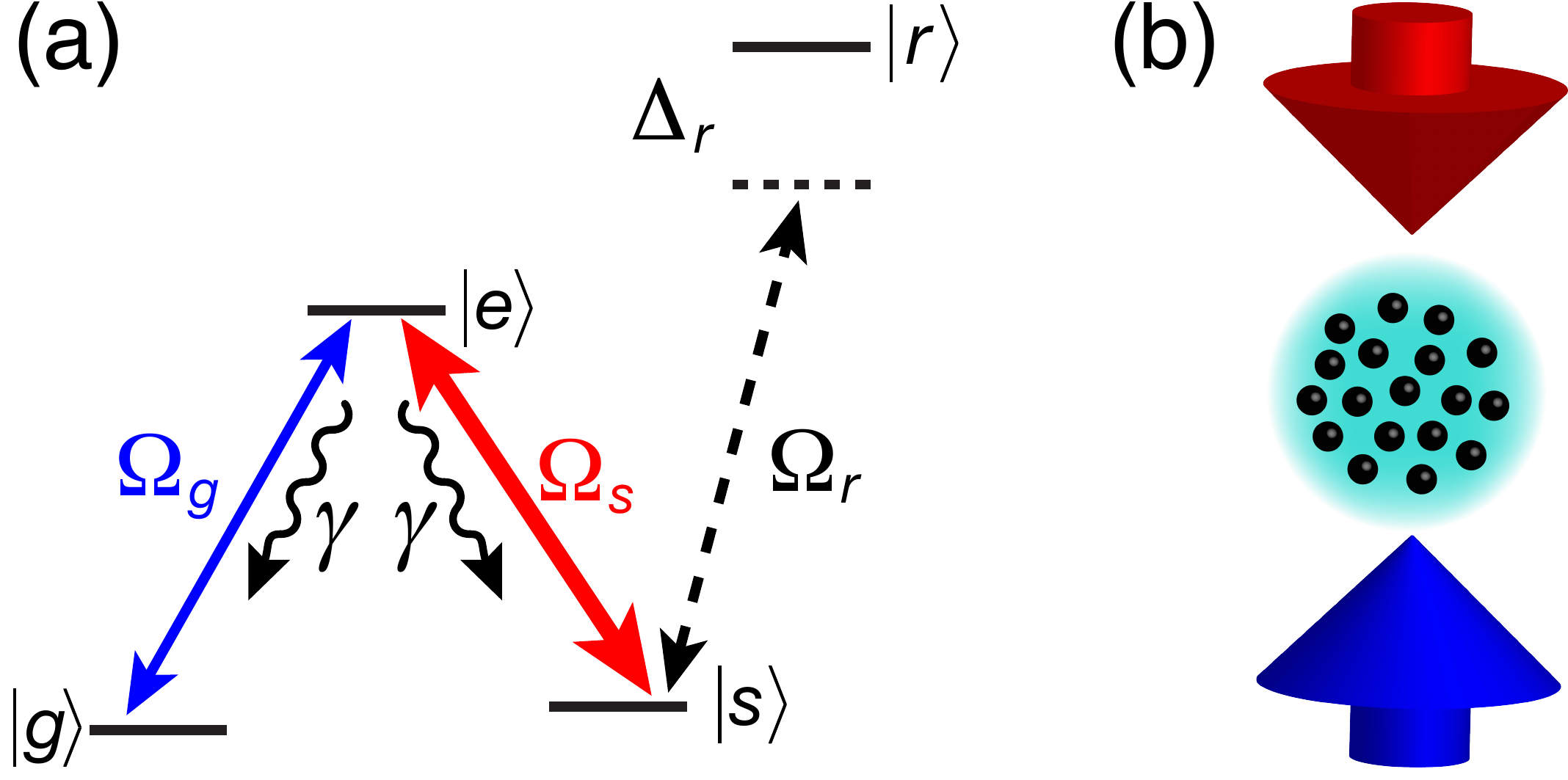}
  \caption{\label{fig:levels}(Color online) (a) Two metastable components of the ground electronic state, $\vert g \rangle$ and  $\vert s \rangle$, are coupled to the electronically  excited state, $\vert e \rangle$, decaying at rate $\gamma$.  State $\vert s \rangle$ is   coupled to a highly-excited Rydberg level, $\ket{r}$, using a far-detuned field,  $\Delta_r \gg \Omega_r$. This results in the soft-core interaction between the atoms in the $\ket{s}$ states, see Fig.~\ref{fig:disp_pots}(a), while states $\vert g \rangle$ and $\vert e \rangle$ are non-interacting. (b) Schematics of the setup: BEC of Rydberg-dressed atoms is confined in a three-dimensional optical dipole trap, fields $\Omega_{g,s}$ are counter-propagating.}
\end{figure}

We consider a Bose-Einstein Condensate (BEC) of atoms with internal $\Lambda$-type linkage pattern and off-resonant Rydberg-dressing, see Fig. \ref{fig:levels}(a). The atoms are initially prepared in the ground state $\vert g \rangle$; the meta-stable state $\ket{s}$ can be chosen as a different fine or hyperfine component of the ground-state manifold. States $\vert g \rangle$ and $\vert s \rangle$ are coupled to an excited state, $\vert e \rangle$, by  laser fields with Rabi-frequencies $\Omega_g$ and $\Omega_s$, respectively. In alkali atoms, $\Omega_{g,s}$ could drive, e.g., the laser-cooling transition, $^2S_{1/2} \leftrightarrow {}^2P_{3/2}$. For simplicity we assume that state $\vert e \rangle$ decays to $\vert g \rangle$ and  $\vert s \rangle$ with the same spontaneous emission rate $\gamma$.

\begin{figure}[br]
\includegraphics[width=0.95\linewidth]{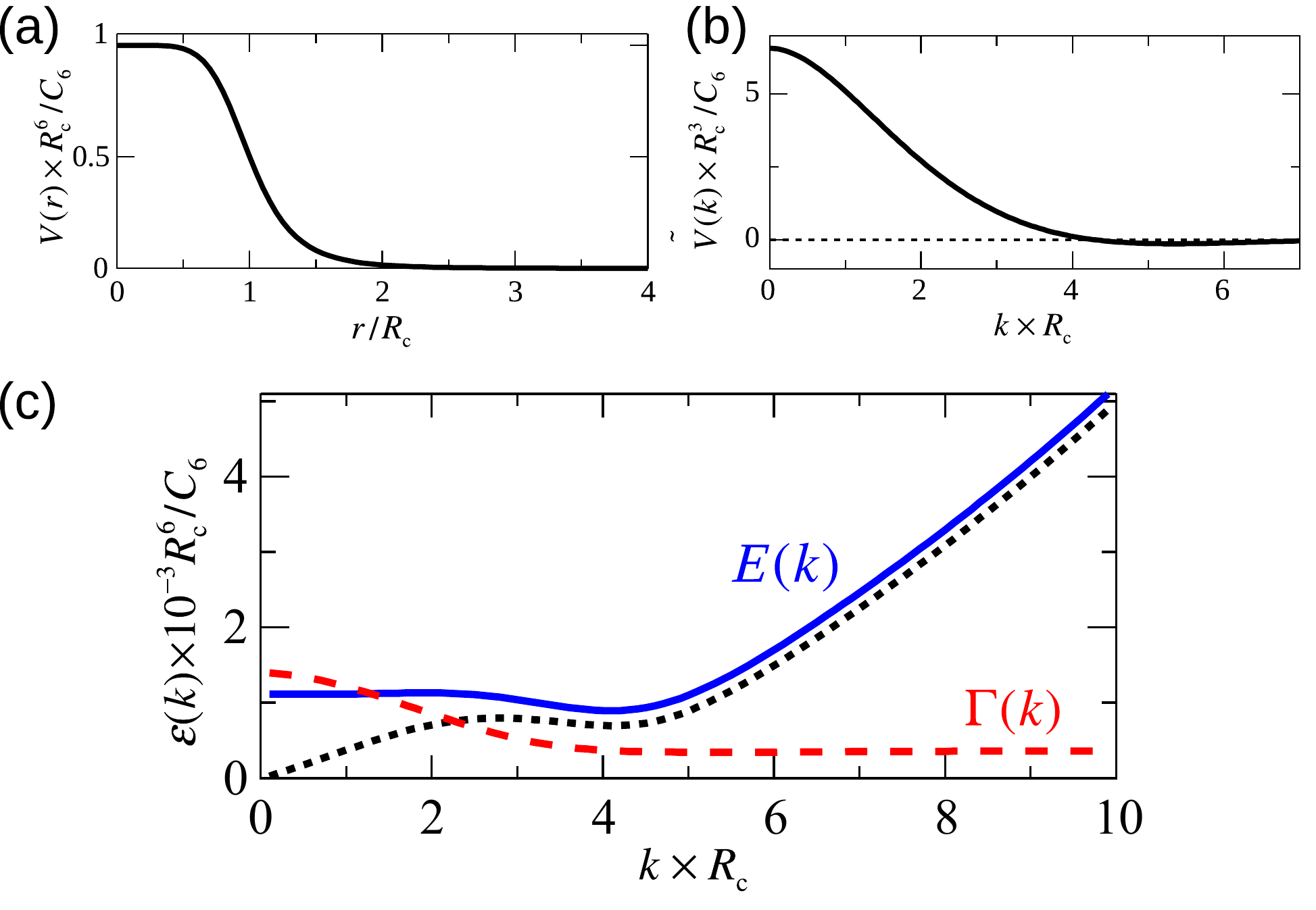}
  \caption{\label{fig:disp_pots}(Color online) (a) Soft-core interaction potential, $V(r)$, between two Rydberg-dressed atoms separated by the distance $r$, and (b) its Fourier transform, $\tilde{V}(k)$. (c) The real part $E(k)$ (blue), and the imaginary part $\Gamma(k)$ (red), of the dispersion relation $\varepsilon(k)$, eq.~(\ref{Epsilon}), possessing a roton minimum, obtained using the self-consistent HFB theory. The black dotted line shows $\varepsilon(k)$ in the absence of dissipation. Parameters correspond to Fig.~\ref{fig:Results}(a).}
\end{figure}

To introduce interactions into the system, we couple the state $\vert s \rangle$  to a high-lying Rydberg state, $\ket{r}$, via a field $\Omega_r$ \cite{Johnson2010,Henkel2010,Pupillo2010, Honer2010}; states $\vert g \rangle$ and $\vert e \rangle$ are not dressed and therefore are non-interacting. In the regime of a far-off-resonant coupling, i.e. for $\Delta_r \gg \Omega_r$, where $\Delta_r=\omega_{r}-\omega_s-\omega_{\Omega_r}$, state $\ket{r}$ can be adiabatically eliminated. Here $\omega_r$, $\omega_s$ and $\omega_{\Omega_r}$ denote the energies of the atomic levels $\ket{r}$, $\ket{s}$ and the carrier frequency of the field $\Omega_r$, respectively. This results in an effective soft-core interaction between the atoms in the $\ket{s}$ state, $V(r)=C_6/(r^6 + R_c^6)$. Here $C_6 = \left(\Omega_r/2\Delta_r \right)^4 C_\text{vdW}$, the cutoff radius $R_c = \left (C_\text{vdW}/2\hbar|\Delta_r| \right)^{1/6}$, and $C_\text{vdW}$ gives the strength of the van-der-Waals interaction in the Rydberg state.
 While the lifetimes of the Rydberg states are on the order of milliseconds, the effective decay rate of the dressed states is given by $\gamma_\text{eff} = \gamma \Omega_r^2/(2 \Delta_r)^2$, which results in the lifetimes as large as several seconds \cite{Henkel2010}. The behavior of $V(r)$ is shown in Fig.~\ref{fig:disp_pots}(a): one can see that at short distances, $r<R_c$, the non-local repulsive tail  $\sim r^{-6}$ turns into a plateau. The competition between the non-local interaction and kinetic energy leads to the formation of a roton minimum as discussed for, e.g., Rydberg-dressed atoms in a BEC \cite{Henkel2010}. It should be noted that the following discussion remains valid if one works with dipolar interactions instead of the screened Rydberg interactions. We focus on a BEC trapped in a three-dimensional optical dipole trap with fields $\Omega_g$ and $\Omega_s$ counter-propagating, as shown in Fig. \ref{fig:levels}(b).

Due to the coupling of the atoms to the electromagnetic fields, the system is subject to spontaneous emission and hence represents an open quantum system. In the most general case, this kind of driven-dissipative dynamics is described by a quantum master equation for the density operator \cite{Breuer2002}. However, in this work we limit ourselves to the regime of weak driving, $|\Omega_{g,s}|^2/\gamma^2 \ll 1$, leading to weak dissipation. This allows us to obtain the dynamics from the solutions of an effective, non-Hermitian Hamiltonian \cite{DalibardPRL92, MolmerJOSAB93, Duerr2009}: $H = H_\text{at} +  H_\text{at-field} +  H_\text{int}$.

Working in a frame rotating with the optical frequencies, the kinetic energy and spontaneous emission of the atoms are given by \cite{Duerr2009} ($\hbar =1$),
\begin{equation}
\label{Hat}
H_\text{at} = \sum_{i,k}\frac{k^2}{2m} \Bd_{i,k}\B_{i,k}   -  i \gamma \sum_{k}  \Bd_{e,k}\B_{e,k}.
\end{equation}
where $\Bd_{i,k}$ ($\B_{i,k}$) is the creation (annihilation) operator for the bosonic field with momentum $k$ and internal state $i \in \{g,s,e\}$, and $m$ is the atomic mass. Neglecting the jump terms corresponds to post-selection of trajectories where no jump occurred. If need be the quantum jumps can be reinstated using the fluctuation dissipation theorem \cite{Louisell-Book}. The interaction with the control fields is given by
\begin{align}
\label{HatField}
H_\text{at-field} =\sum_k &  \left( \Omega_g \Bd_{e,k+q_g}\B_{g,k} + \Omega_s \Bd_{e,k+q_s} \B_{s,k}+ \text{h.c.} \right)\nonumber \\
& +\delta\Bd_{s,k}\B_{s,k}+\Delta\Bd_{g,k}\B_{g,k},
\end{align}
where $q_g$ and $q_s$ are the wave-vectors of the corresponding laser fields with Rabi frequencies $\Omega_g$ and $\Omega_s$. The one- and two-photon detunings are given by $\Delta = \omega_{e}-\omega_g-\omega_{\Omega_g}$ and $\delta = \omega_{s}-\omega_g+\omega_{\Omega_g}-\omega_{\Omega_s}$, respectively. Finally, the interaction between the $\vert s \rangle$-states reads
\begin{equation}
\label{Hdd}
H_\text{int}=\frac{1}{2}\sum_{k,k',q} v_q \Bd_{s,k+q}\Bd_{s,k'-q}\B_{s,k'}\B_{s,k}.
\end{equation}
Here $v_q= (2\pi)^3/V \times \tilde{V}(q)$, where $V$ is the volume of the BEC and $\tilde{V}(q)$ is a 3D Fourier transform of the soft-core interaction $V(r)$, given by  $ \tilde{V}(\zeta) = \frac{2\pi^2 C_6}{3 R_c^3 \zeta} \e^{- \zeta/2} \left[\e^{-\zeta/2}-\cos(\sqrt{3}\zeta/2) + \sqrt{3}\sin(\sqrt{3}\zeta/2) \right]$, with $\zeta = R_c |q|$ (cf. Fig. \ref{fig:disp_pots}(b)). The presence of a harmonic trap does not significantly alter the shape of interactions between Rydberg-dressed atoms~\cite{Macri13}. Furthermore, we assume that the size of the trap substantially exceeds the length scales of interest, which allows us to neglect the coupling of atomic motion to the trapping potential.

In the absence of interactions, the Hamiltonian $H$ exhibits a momentum dependent dark-state $\ket{D,k}\sim \Omega_s\ket{g, k-q_g}-\Omega_g \ket{s, k-q_s}$, which decouples from all other states if $k = (-2\delta m + q_g^2-q_s^2)/(2(q_g-q_s))$. As a consequence this dark-state is insensitive to spontaneous decay and hence long-lived \cite{Bergmann-RMP-1998}. The existence of this dark-state constitutes the foundation of velocity selective coherent population trapping (VSCPT) \cite{Aspect1988}. The effect of the interactions is equivalent to a momentum- and density-dependent two-photon detuning $\delta(k)$~\cite{Otterbach2013}. Thus, the existence and the momentum of the dark-state depend on the many-body properties of the system.

In order to make the above arguments more quantitative, we define dark- and bright-states as $\D_k = \cos\theta\B_{g,k}-\sin\theta\B_{s,k+q_g-q_s}$ and $\hat\psi_k = \sin\theta\B_{g,k}+\cos\theta\B_{s,k+q_g-q_s}$, where the mixing angle is given by $\tan\theta = \Omega_g/\Omega_s$ and $\Omega^2 = \Omega_g^2+\Omega_s^2$. The states $\ket{g}$ and $\ket{s}$ can be chosen as two hyperfine components of the electronic ground state, such that the net momentum transfer, $\Delta q = q_g - q_s=(\omega_{\Omega_g}-\omega_{\Omega_s})/c\approx 0$, can be neglected. Furthermore, we set $\Delta=\delta=0$ in order to avoid interaction-induced resonances, i.e. points where the bare detuning and the interaction cancel each other. Writing the full Hamiltonian $H$ in terms of dark- and bright-states, we derive an effective Hamiltonian for the weakly decaying dark-states using perturbation theory. Under the assumption $\gamma \gg \Omega, v_0$, we first adiabatically eliminate the excited states. This results in 
an effective Hamiltonian for the dark- and bright-states:
\begin{align}
 H_0 = \sum_k \left(\frac{k^2}{2m} -i\frac{\Omega^2}{\gamma}\right)\hat\psi_k^\dagger\hat\psi_k +\frac{k^2}{2m} \D_k^\dagger\D_k
\end{align}
The interactions are incorporated by assuming the BEC to be weakly interacting, i.e. $a_s (N/V)^{1/3}<1$, where the $s$-wave scattering length of the dark-states is given by $a_s = (m C_6)^{1/4} \sin \theta$. This allows us to treat the quickly decaying bright-states to be in vacuum and hence we need only keep terms up to first order in $\hat\psi_k$, leading to
\begin{align}
 H_{\rm int} =& \frac{\sin^4\theta}{2}\sum_{k,k',q} v_q \Dd_{k+q}\Dd_{k'-q}\D_{k'}\D_{k} \\
 &+\frac{\sin^3\theta\cos\theta}{2}\sum_{k,k',q} v_q \Dd_{k+q}\left(\Dd_{k'-q}\hat\psi_{k'}+{\rm h.c.}\right)\D_{k} \nonumber
\end{align}

In order to solve the many-body problem described by the Hamiltonian $H=H_0 + H_\text{int}$, we first apply the Bogoliubov approximation $\D_0 = \langle\D_0\rangle \equiv N_0$ and $\hat\psi_0 =\langle \hat\psi_0\rangle= 0$ and subsequently eliminate the bright-states assuming $\Omega^2/\gamma \gg v_0, \langle k^2/(2m)\rangle$. 
To account for the far-from-equilibrium behavior of the system, we work within the Hartree-Fock-Bogoliubov (HFB) approximation~\cite{sup, MazetsJPB04, GriffinPRB96}, taking into account higher-order correlations. This allows us to include the interactions between different momentum modes and the resulting redistribution of populations during the dissipative evolution, absent in the standard Bogoliubov approximation. Applying the Popov approximation \cite{sup, PopovBook, GriffinPRB96} results in a gapless spectrum of the fundamental excitations and allows a proper treatment of the thermal condensate fraction. After deriving the HFB equations of motion and performing the Bogoliubov transformation, $\D_k\,=\,\alpha_k\be_k+\beta_{-k}^*\bed_{-k}$, we obtain the low-energy quasiparticle spectrum:
\begin{equation}
\label{Epsilon}
\varepsilon^2(k) =  \xi(k)  \left[\xi(k) + 2\sin^4\theta g_k - i  \varkappa 4 g_ k (g_0 + g_k)\right].
\end{equation}
Here $\xi(k) =  k^2/(2m) + [W(k) - W(0)]\sin^4 \theta - i\varkappa g_0^2$; the interaction strength and dissipation rate parametrized respectively by $g_k = v_k N_0 \sim C_6 \rho/R_c^3$ and $\varkappa= (\gamma/\Omega^2) \sin^6 \theta \cos^2 \theta$. The correlation term $W(k) = \sum_{\mathbf{q \neq k}} v_{q} \langle \Dd_{|\mathbf{k+q}|} \D_{|\mathbf{k+q}|} \rangle$ is a result of the HFB approach and accounts for the interaction between different $k$-modes and has to be evaluated self-consistently (see~\cite{sup}). Neglecting $W(k)$ leads to the standard Bogoliubov results.

\begin{figure}[br]
\includegraphics[width=1\linewidth]{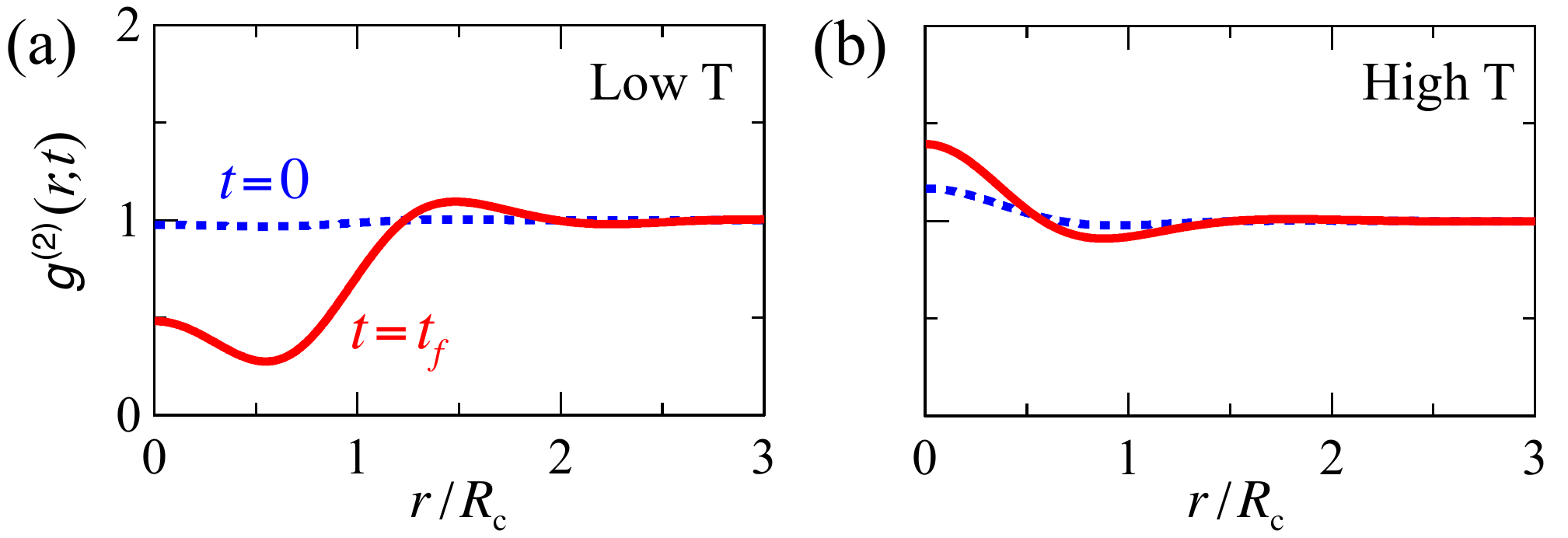}
  \caption{\label{fig:Results}(Color online) Time evolution of the pair correlation function, eq.~(\ref{PairCorrFunc}), from time $t=0$ (blue dotted lines) to $t=t_f$ (red solid lines), obtained within the self-consistent HFB approach. The initial BEC temperatures are: (a) $T \approx 0.05T_c$, (b) $T \approx 0.4T_c$. The results correspond to the density $\rho = 10^{13}~\text{cm}^{-3} = 10 R_c^{-3}$, the interaction constant $C_6 = 5 \times 10^{9}$~a.u., and the Rabi frequency $\Omega=\gamma/10$, with $\gamma = 2\pi \times 5.2$~MHz (the $6^2S_{1/2} \leftrightarrow 6{}^2P_{3/2}$ transition of a Cs atom). The mixing angles are $\theta=0.95$ (a) and $\theta=0.65$ (b).}
\end{figure}

The complex relation (\ref{Epsilon}) can be represented as $\varepsilon(k) = E(k) - i\Gamma(k)$. The real part, $E(k)$, gives the quasi-particle dispersion, shown by the blue line in Fig.~\ref{fig:disp_pots}(c). For sufficiently strong interactions, the dispersion curve $E(k)$ shows a roton minimum around $k_{\rm rot} R_c \approx 4.2$ (for the parameters used in the calculation) due to the negative part of the potential $v_k$. The emergence of the roton minimum has already been discussed in a number of works on Rydberg-dressed BEC's \cite{Henkel2010} and dipolar BEC's in confined geometries \cite{SantosPRL03, OdellPRL03, GiovanazziEPJD04, LahayePfauRPP2009}. Contrary to these works, we focus on the dissipative dynamics of the system, given by the momentum-dependent imaginary part $\Gamma(k)$, which is shown by the red line in Fig.~\ref{fig:Results}(a). The behavior of $\Gamma(k)$ originates in the shape of the soft-core potential in the momentum space, $v_k$, cf. Fig. \ref{fig:disp_pots}(b). 
Being large at small momenta, the dissipation rate decreases with growing $k$ and features a shallow minimum in the region where $v_k<0$. The position of the roton minimum can be tuned in experiment by changing the cutoff radius $R_c$ and the mixing angle $\theta$, while the driving field $\Omega$ sets the magnitude of the dissipative term $\Gamma(k)$.
 
In order to understand the dissipative formation of a non-local spatial structure, we consider a BEC at low, but finite temperature $T>0$ (although an analogous argument can be made for $T=0$).  The phonon modes are then populated according to the Bose-Einstein distribution, $n (k, T) = 1/[\e^{E(k)/k_BT}-1]$. In order to maximize the visibility of the non-local correlations, we adjust the parameters of the dissipation such that the decay rate near the roton minimum is smaller than for the rest of the populated phonon modes with $k<k_\text{rot}$. This ensures that the roton states, whose thermal population is already larger than for the rest of the modes, also decay at a slower rate (cf. Fig. \ref{fig:disp_pots}(c)). We note that   the presence of the roton minimum  facilitates the preparation of states with a narrow distribution of momenta. During the time evolution, the phonon modes undergo photon-assisted dissipation at rate $\Gamma (k)$, resulting in a dissipative shaping of the momentum-distribution or 
in population decay due to the trap loss. Since $\Gamma (k)$ is small around $k=k_\text{rot}$, the population in the vicinity of the roton minimum decays slower than that of other momentum classes. The peak in the momentum distribution emerging after the evolution time corresponds to density oscillations in real space, with wavelength $\lambda_\text{osc} \sim 2\pi/k_\text{rot}\approx 1.5 R_c$, in good agreement with our numerical findings and signaling the onset of  an ordered structure.

To study the dynamics of the ordering, we evaluate the structure factor which becomes explicitly time-dependent due to dissipation~\cite{sup}:
\begin{equation}
\label{StrFact}
 S(k,t) \,=\,   \left \vert \frac{  \xi(k)  }{\varepsilon(k)}   \right \vert
\left [ 2 \langle \bed_k (t) \be_{k} (t) \rangle  +1 \right],
\end{equation}

In the limit of $\Gamma(k) \to 0$ and $T\to0$, eq. (\ref{StrFact}) coincides with the Feynman relation $S(k,0) = k^2/[2 m E(k)]$.  To guarantee the validity of our theory, we focus on the regime of small thermal excitation, i.e. $N_e \equiv \sum_{\mathbf{k}\neq 0} \langle \bed_{\mathbf{k}} \be_{\mathbf{k}} \rangle \ll N_0$. As we are operating an open system whose true steady-state is the vacuum, there is a natural upper limit, $t_f$, for the time-scale on which our theory is valid. This timescale is given by the breakdown of the HFB theory when $N_e(t_f) \approx N_0(t_f)$ or $T_c (t_f) \approx T$, where $T_c$ is the density-dependent critical temperature of the BEC. The breaking of translational symmetry manifests itself as oscillations in the pair correlation function, $g^{(2)}(r,t)$, which is given by the Fourier transform of the structure factor \cite{Pitaevskii2003}:
\begin{equation}
\label{PairCorrFunc}
g^{(2)} (r,t) = 1 + \frac{1}{(2\pi)^3 \rho (t)} \int  d \mathbf{k} \left[ S(k,t) - 1 \right] e^{-i \mathbf{k \cdot r}},
\end{equation}
where $\rho(t)$ is the now time-dependent particle density. A peak in $S(k)$, corresponding to a minimum in $\varepsilon(k)$, hence leads to an oscillatory behavior with the wave-length determined above.

Fig. \ref{fig:Results}(a) shows the time evolution of the pair correlation function for the condensate with initial temperature $T\approx 0.05 T_c (0)$, where  $T_c = 56$~nK for the parameters used in Fig.~\ref{fig:Results}. To obtain the far-from-equilibrium dynamics including population-redistribution and interactions between different momentum modes, the HFB equations were solved self-consistently evaluated at every time step~\cite{sup}. At time $t=0$ the pair correlation function is close to unity (blue dashed line), showing the absence of non-local correlations. During the time evolution, the driven-dissipative dynamics results in pronounced oscillations in $g^{(2)} (r,t)$ (red solid line). Due to the dissipation, the density of the BEC decreases by about a factor of $\sim$20 due to the loss of atoms from the trap, which results in $T\approx 0.4 T_c (t_f)$. For the parameters used in Fig. \ref{fig:Results}, the evolution time $t_f \sim 350~\mu$s. A fingerprint of the onsetting non-local order that can be most easily detected is the anti-bunching of $g^{(2)}(r)$ at $r=0$~\cite{Solyom2007, GorshkovPRL11, Peyronel2012}.   However, in order to clearly demonstrate the  non-local correlations, the full off-diagonal pair-correlation function needs to be measured. This 
can be done using, e.g., Bragg scattering~\cite{Stamper-Kurn1999}, noise correlation \cite{Altman2004,Folling2005},  or time-of-flight~\cite{JonaLasinioPRA13}  spectroscopy.

If the initial condensate temperature is higher, such that $g^{(2)}(0,0)$ is bunched, the presented approach still leads to similar results. Fig. \ref{fig:Results}(b) shows $g^{(2)}(r,t)$ evolving from the initial temperature $T\approx 0.4 T_c (0)$. In this case the interatomic interactions were decreased by appropriately tuning the mixing angle $\theta$, in order to avoid the condensate depletion due to the interactions between populated high-momentum modes. As one can see, at high temperatures the evolution time of $t_f \sim 400 \mu$s suffices for formation of visible oscillations in $g^{(2)}(r,t_f)$ that can be readily detected experimentally. To understand the crossover between the enhancement of anti-bunching and bunching, we note that higher temperatures favor more high-energy fluctuations. However, the dissipation is minimal around a certain energy range. Hence at sufficiently high temperatures the thermal fluctuations are not damped but enhanced due to the loss of small energy excitations, leading to bosonic bunching. Bunching and anti-bunching in weakly interacting BECs have been observed in the temporal domain in Refs.~\cite{Guarrera2011,
Guarrera2012}. We note that due to the dissipative nature of the non-local order, long coherence times of Rydberg dressing are not required to observe it, as opposed to proposals based on coherent evolution. The necessary coherence times longer than the lifetimes of the dark states, $t_f \sim 1$~ms, can be achieved in current experiments \cite{Schempp2010,Pritchard2010,Nipper2012,Dudin2012,Peyronel2012,SchaussNat12, Low2012}.

 
In summary, we presented a technique for the dissipative preparation of non-local ordered structures in a BEC of Rydberg-dressed atoms. We showed that using laser-induced dressing, one can engineer momentum-dependent dissipation that exhibits an almost dark phonon mode at a nonzero momentum, resulting in non-local spatial correlations. The range of the correlations is adjustable by tuning the control fields $\Omega_r$ and $\Delta_r$, as well as the mixing angle $\theta$. Additional control can be introduced using the bare two-photon detuning, and thus exploiting interaction induced resonances.  Dissipatively  ordered structures can be created and studied in current experiments with ultracold atoms.  While the presence of harmonic confinement can alter the phonon spectrum~\cite{BissetPRA13, JonaLasinioPRA13b}, compared to `flat traps'~\cite{GauntPRL13}, the dissipation-induced effects are expected to hold in confined geometries. Although we exemplified the technique with Rydberg-dressed atoms, a similar behavior can 
be observed in dipolar gases confined to one- and two-dimensional optical traps \cite{SantosPRL03, OdellPRL03, GiovanazziEPJD04, LahayePfauRPP2009}. This allows to apply the technique to strongly magnetic atoms \cite{LuPRL11} or laser-cooled molecules~\cite{ShumanNature10} dressed with microwave fields~\cite{LemeshkoPRL12}. Extensions to interactions with different symmetries, such as electric quadrupole-quadrupole ones \cite{BhongalePRL13}, also seem possible.

Both authors contributed equally to this work. We thank H. Weimer, T. Pohl, J. Bohn, and R. Schmidt for insightful discussions. This work was supported by NSF through a grant for the Institute for Theoretical Atomic, Molecular, and Optical Physics at Harvard University and Smithsonian Astrophysical Observatory as well as the Harvard Quantum Optics Center.

\bibliography{References}

\end{document}